\documentclass[twocolumn]{revtex4}
\usepackage{graphicx}
\usepackage{dcolumn}
\usepackage{longtable}
\usepackage{amsmath}
\usepackage{amssymb}

\begin{document}

\title
{Nuclear shape transitions, level density, and underlying interactions}

\author
{S. Karampagia and Vladimir Zelevinsky}

\affiliation
{National Superconducting Cyclotron Laboratory and Department of Physics and Astronomy, Michigan State University, East Lansing, MI 48824-1321, USA}

\begin{abstract}

{\bf Background}: The configuration interaction approach to nuclear structure uses the effective Hamiltonian in a finite orbital
                  space. The various parts of this Hamiltonian and their interplay are responsible for
                  specific features of physics including the shape of the mean field and level density.
                  This interrelation is not sufficiently understood.                 \\
                         \\
{\bf Purpose}: We intend to study phase transitions between spherical and deformed shapes
               driven by different parts of the nuclear Hamiltonian and to establish the presence of the   collective enhancement of
               the  nuclear level density by varying the shell-model matrix elements. \\
               \\

{\bf Method}: Varying the
              interaction matrix elements we define, for nuclei in the $sd$ and $pf$ shells,
              the sectors with spherical and deformed shapes. Using the moments method that does
              not require the full diagonalization we relate
              the shape transitions with the corresponding level density. \\
              \\

{\bf Results}: Enhancement of the level density in the low-energy part of the spectrum is
               observed in clear correlation with a deformation phase transition induced mainly by the
               matrix elements of single-particle transfer. \\
               \\

{\bf Conclusions}: The single-particle transfer matrix elements in the shell model nuclear Hamiltonian
                   are indeed the carriers of deformation, providing
                   rotational observables  and enhanced level densities.

\end{abstract}

%\pacs{21.60.Cs, 21.10.Ma, 05.30.Rt}

\maketitle

\section{Introduction}

The knowledge of the level density in a quantum many-body system is necessary
for the correct understanding of the response of the system to external perturbations.
The nuclear level density is a vitally important element of reaction theory,
including astrophysical processes and broad applications of nuclear physics.
But it might also serve as a mirror reflecting special features of intrinsic structure
and this will be the main subject of our consideration.

In a Fermi-system environment, the level density exponentially grows with energy
due to the combinatorics of particle-hole excitations from the defrosted
Fermi surface. This occurs even in the simplest picture of a Fermi gas without
residual interaction \cite{Fermi1,Fermi2}.  The Fermi gas
model does not however account for the effects on the level density
due to the shell structure, pairing correlations
\cite{pairing1,pairing2} or coherent excitations of collective nature
\cite{collect1, collect2}. Various semi-phenomenological
approaches have been developed which account for such
effects
\cite{Ignatyuk,Dossing} considered as additions to the
skeleton of the Fermi-gas, or of a more elaborate self-consistent mean field with pairing.

Low-lying collective modes, mainly of isoscalar nature, lead to the reconfiguration of
the nuclear spectra. In an even-even non-magic nucleus, pairing correlations create
an energy gap in the excitation spectrum. Inside the gap vibrational collective modes
start the sequence of phonon states which gradually mix with unpaired particles appearing
above the pair breaking threshold. Away from the magic nuclei,
the accumulating valence particle frequently lead to broken internal
symmetry and static deformation of the core.
Then nuclear rotation appears as a new branch of the excitation spectrum.
The rotational bands, with a small distortion of the nuclear field along the band, appear at low energy.
All these effects should noticeably change the low-lying nuclear level density
\cite{Bjornholm, Bohr2, Ericson}.

In the framework of the shell model, pairing and collective effects are fully taken
into account through the two-body interaction matrix elements. Since the shell model
is formulated in a truncated orbital space and therefore has the fixed total
number of quantum states, the collective enhancement can appear as enrichment of the
level density at the low-energy part of the spectrum, accompanied by a corresponding
suppression of the level density at higher excitation energy. The shell-model
experience shows that the effects of deformation and related rotational motion appear
naturally for a sufficiently rich space and appropriate set of two-body interaction
matrix elements as a result of the diagonalization in a spherical basis. This is an
important advantage of the shell-model approach since one does not need to take special
care of the strict fulfillment of conservation laws (particle number, angular momentum,
parity and isospin). On the other hand, the computational problems impose the limitation 
on the total dimension of the orbital space.

The {\sl moments method} based on statistical properties of large Hamiltonian matrices
\cite{kota,wong} was recently formulated \cite{PLB11,CPC13} as a practical tool for
calculating the level density for a given Hamiltonian avoiding the diagonalization
of large matrices. It was shown how the first two moments of the Hamiltonian define
the full level density that coincides with the result of the exact diagonalization if
the latter is feasible. Some latest results and first comparisons with the phenomenology,
thermodynamics, and mean field combinatorics can be found in \cite{SZ15}. One important
conclusion is that in realistic cases the level density in a finite Hilbert space is a
smooth bell-shape curve. The contributions of individual shells, which are clearly
pronounced in the mean-field combinatorics, are smeared by the multitude of incoherent
collision-like interactions which are always present in realistic Hamiltonians in addition
to the collective parts like pairing and multipole forces.
In this article, our problem is rather different. We are going to explore the landscape
of nuclear Hamiltonians varying the parameters in order to establish the existence and
dynamic sources of the collective enhancement of the level density. We use the moments method
to extract the cases with collective behavior 
and study the corresponding level densities. This method was used earlier \cite{Horoi} 
for understanding the predominance of prolate deformation among non-spherical nuclei. 
With the variation of parameters of the shell-model Hamitonian, we are able to 
localize and study the phase transitions between spherical and deformed shapes. 

In Section 2 we give a  brief description of the moments method, in Section 3 we describe the 
division of the full shell-model Hamiltonian into  different subsets of matrix elements which 
can be varied independently.  Section 4 presents  the effects of those subsets on 
the low-lying spectrum of different nuclear systems and on the level densities. In 
Section 5 we discuss a quantum phase transition between spherical and deformed shapes by 
varying the strength of the matrix elements. The concluding discussion is given in Section 6.

\section{Moments method}

Here we very briefly remind the formalism of the moments method. We use the shell-model Hamiltonian
$H$ that contains the mean field and residual two-body effective interactions. The level density
is found as a superposition of modified (finite range) Gaussians,
\begin{equation}
\rho(E;\alpha)=\sum_{p}D_{\alpha p}G_{\alpha p}(E).                          \label{1}
\end{equation}
Here $\alpha$ stands for the exact quantum numbers of spin and parity, while $p$ runs over partitions
(distributions of particles among available single-particle orbitals); $D_{\alpha p}$ is the dimension
of a given partition, and $G_{\alpha p}$ is the finite-range Gaussian determined by the
ground state energy, the centroid (the first moment of the Hamiltonian) and the width (the second moment).
The second moment includes all interactions mixing the partitions.
Both moments can be computed directly by the Hamiltonian matrix avoiding its diagonalization. As we have
already mentioned, the result (\ref{1}) is, in all studied cases, in good agreement with the product
of the full diagonalization if the latter is practically possible.

Technical details related to finding the ground state energy, $M$-scheme against
$J$-scheme, fit of the spin cut-off parameter, removal of unphysical center-of-mass excitations in the
cases of cross-shell transitions etc. are discussed in previous publications
\cite{PLB11,CPC13,SZ15}.

\section{Searching for collective effects}

In the simplest (but still rich in physics and numerous applications) case of
the $sd$ shell model we have only three single-particle levels, $1s_{1/2},0d_{5/2},0d_{3/2}$.
The angular momentum and isospin conservation allow 63 matrix elements of the residual
two-body interactions. Keeping intact all symmetry requirements, we can vary
numerical values of the two-body matrix elements of the effective interaction.
As a result, we come to different versions of the shell model which can cover the whole
spectral variety allowed by the given Hilbert space. In this way we can select the parts
of the interaction responsible for specific observable physical phenomena.

In a recent study \cite{Horoi}, where in the same spirit the $pf$ orbital space
was used, it was found that certain interaction matrix elements are responsible
for the transition from a spherical shape to a deformed one. First of all, they
were the matrix elements ($pf$ matrix elements in that specific model) changing 
the occupation numbers of the subshells by one unit, i.e. the matrix elements 
$\left< j_k,j_l|V|j_m, j_n \right>$ with $j_k=j_m$, or $j_k=j_n$, or $j_l=j_m$, 
or $j_l=j_n$. This drives the mixing of spherical orbitals in the process of
deformation. A complementary version of a similar approach was applied 
in \cite{SZ15}
in order to demonstrate that the incoherent parts of the residual interaction are
essential for producing chaotic wave functions and resulting smooth level density.

Borrowing this approach we divide the set of interaction matrix elements into two parts.
The part $V_{1}$ includes the ``particle-hole" matrix elements which change 
the occupation number of the subshells by one unit with the change of orbital momentum 
$\Delta\ell=0\,{\rm or}\,2$, whereas the part $V_{2}$ includes the remaining matrix elements, 
which either don't change the occupation number of the subshells 
($j_k=j_m$ and $j_l=j_n$), or change it by two units ($j_k \neq j_m$ and $j_k \neq j_n$),
\begin{equation}
H=h +  k_1 V_{1} +k_2 V_{2};                                   \label{2}
\end{equation}
here the part $h$ contains the single-particle energies.
From this point on we will be calling the matrix elements of the $V_{1}$ part, 
``one unit change" matrix elements.
The numerical parameters $k_{1}$ and $k_{2}$ allow us to explore regions of 
the Hilbert space where the nuclear structure undergoes significant changes.
The original shell-model case emerges for $k_1=k_2=1$.
Probing various combinations of parameters $k_{1}$ and $k_{2}$ one can see how these
two parts affect the level density and other observable quantities of interest.
We will study the evolution of the level density as a function of these
particular interaction modes paying special attention to the low-lying parts of the
spectrum as indicators of characteristic underlying structures. In even-even nuclei we
characterize the low-lying spectrum by the levels
($2^+$, $4^+$, $6^+$), quadrupole transitions between them, and shape multipoles, as well
as by the resulting level density.

\section{Exploring the nuclear landscape}

This section presents a quantitative study of how the $V_1$ and $V_2$ parts
of the shell-model Hamiltonian (\ref{2}) change the collective observables. We find whether
these interactions are capable of producing typical deformed or
spherical characteristics of the nuclear field in the low-energy part of the nuclear spectrum.

Two shell model spaces, the $sd$ and $pf$, have been studied. As already mentioned,
the interaction of the $sd$ shell-model space has 63 non-zero matrix elements, among 
which 22 elements induce one-body transitions between the partitions (these are  included in the
$V_1$ part of eq. (\ref{2}), while the remaining  41 matrix elements (those included in 
the part $V_2$  of eq. (\ref{2}) either couple states within the same partition or 
transfer two particles between partitions, including usual pairing.

In the same spirit, the interaction in the $pf$ shell has 195 non-zero matrix
elements, 79 of which are included in the $V_1$ part of eq. (\ref{2}), while 116 remaining
matrix elements, which either don't change the occupation number of the subshells or induce 
two-body transitions between the partitions, make up the $V_2$ part of eq. (\ref{2}). We 
have considered the cases with four valence protons + four valence neutrons and six valence 
protons + six valence neutrons for the $sd$ shell (these correspond to the $^{24}$Mg and 
$^{28}$Si nuclei, respectively) and the case of six valence protons + six valence neutrons 
for the $pf$ shell model (the $^{52}$Fe nucleus).

The observables used for studying the effects of various parts of the interaction are the
low-lying $2^+_1$ and $4^+_1$ energy levels, the ratio of these energies, 
$R_{4/2}$=$E(4^+_1)/E(2^+_1$), the expectation value of the quadrupole moment in the first 
excited state, $Q(2^+_1$) and the reduced quadrupole transition probability, 
$B$(E2;$2^+_1\rightarrow 0^+_1$). In order to distinguish  between spherical 
and deformed cases, we use as an indicator the ratio $R_{4/2}$, which should be 
close to 2 for spherical shapes and close to 3.3 for deformed shapes. Selected 
results of the exact shell-model analysis are shown in Tables I and II.

\begin{table}
\caption{Results for $k_1$=1.0 and changing $k_2$ for yrast energies (MeV) of $2^+$ and $4^+$ levels, ratios $R_{4/2}$,
quadrupole moments $Q(2^+_1)$ (e$\cdot$fm$^2$) and reduced transition probabilities 
$B$(E2;2$^+_1 \rightarrow$ 0$^+_1$) (${\rm e}^2\cdot{\rm fm}^4$) for $^{28}$Si, $^{24}$Mg, and $^{52}$Fe. }\label{SiMgFe_k1}
\bigskip
\begin{center}
\begin{tabular}{ r  r  r  r  r  r  r }
\hline
\hline
   $^{28}$Si,    & $k_1$=1.0  &   &   &  &  \\
     $k_2$       & $E$(2$^+_1$) & $E$(4$^+_1$) & $R_{4/2}$ & $Q(2^+_1$) & $B$(E2;2$^+_1 \rightarrow$ 0$^+_1$) \\  \hline

0.0 & 0.964   &  3.197     & 3.32  & -11.50     & 30.75   \\
0.1 & 0.866   &  2.797     & 3.23  & -12.21     & 35.54   \\
0.2 & 0.775   &  1.881     & 2.42  & 20.12       & 9.75   \\
0.3 & 0.628   &  1.966     & 3.13  & 21.06       & 109.4  \\
0.4 & 0.685   &  2.266     & 3.31  & 21.46       & 112.7  \\
0.5 & 0.781   &  2.603     & 3.33  & 21.62       & 113.9  \\
0.6 & 0.925   &  2.975     & 3.21  & 21.61       & 113.0  \\
0.7 & 1.122   &  3.377     & 3.01  & 21.52       & 110.5  \\
0.8 & 1.369   &  3.801     & 2.78  & 21.33       & 107.2  \\
0.9 & 1.659   &  4.233     & 2.55  & 21.27       & 103.6  \\
1.0 & 1.987   &  4.658     & 2.34  & 18.79       & 81.93  \\  \hline \hline
   $^{24}$Mg,    & $k_1$=1.0  &   &   &  &  \\
     $k_2$       & $E(2^+_1)$ & $E(4^+_1)$ & $R_{4/2}$ & $Q(2^+_1$) & $B$(E2;2$^+_1 \rightarrow$ 0$^+_1$) \\  \hline

0.0&  0.596 &  1.667 &  2.80 &  -16.32 &  78.09 \\					
0.1&  0.636 &  1.795 &  2.82 &  -18.04 &  83.40 \\					
0.2&  0.661 &  1.931 &  2.92 &  -18.80 &  86.61 \\					
0.3&  0.689 &  2.095 &  3.04 &  -19.28 &  89.04 \\					
0.4&  0.731 &  2.297 &  3.14 &  -19.59 &  90.99 \\					
0.5&  0.794 &  2.541 &  3.20 &  -19.77 &  92.54 \\					
0.6&  0.882 &  2.828 &  3.21 &  -19.84 &  93.83 \\					
0.7&  0.998 &  3.158 &  3.16 &  -19.81 &  94.87 \\					
0.8&  1.142 &  3.529 &  3.09 &  -19.70 &  95.75 \\					
0.9&  1.313 &  3.937 &  3.00 &  -19.51 &  96.27 \\					
1.0&  1.509 &  4.378 &  2.90 &  -17.44 &  79.12 \\	\hline \hline	
   $^{52}$Fe,    & $k_1$=1.0  &   &   &  &  \\
     $k_2$       & $E(2^+_1$) & $E(4^+_1$) & $R_{4/2}$ & $Q(2^+_1$) & $B$(E2;2$^+_1 \rightarrow$ 0$^+_1$) \\  \hline
0.0 &	0.296 &	0.771 &	2.60 &	-20.68 &	92.28 \\		
0.1 &	0.312 &	0.854 &	2.74 &	-25.76 &	152.60 \\		
0.2 &	0.319 &	0.951 &	2.98 &	-27.16 &	168.80 \\		
0.3 &	0.353 &	1.081 &	3.06 &	-27.92 &	176.70 \\		
0.4 &	0.401 &	1.232 &	3.07 &	-28.48 &	182.40 \\		
0.5 &	0.461 &	1.397 &	3.03 &	-28.92 &	187.50 \\		
0.6 &	0.528 &	1.576 &	2.98 &	-29.32 &	192.60 \\		
0.7 &	0.604 &	1.768 &	2.93 &	-29.70 &	198.20 \\		
0.8 &	0.688 &	1.975 &	2.87 &	-30.07 &	204.70 \\		
0.9 &	0.781 &	2.197 &	2.81 &	-30.42 &	212.20 \\		
1.0 &	0.883 &	2.434 &	2.76 &	-30.76 &	221.10 \\	\hline	\hline
\end{tabular}
\end{center}
\end{table}

\begin{table}

\caption{Results for $k_1$=1.0 and changing $k_2$ for yrast energies (MeV) of $2^+$ and $4^+$ levels, ratios $R_{4/2}$,
quadrupole moments $Q(2^+_1$) (${\rm e}\cdot{\rm  fm}^2$) and reduced transition probabilities $B$(E2;2$^+_1 \rightarrow$ 0$^+_1$) (${\rm e}^2\cdot{\rm fm}^4$) for $^{28}$Si, $^{24}$Mg, and $^{52}$Fe. }\label{SiMgFe_k2}

\bigskip
\begin{center}
\begin{tabular}{ r  r  r  r  r  r  r }
\hline
\hline
   $^{28}$Si,    & $k_2$=1.0  &   &   &  &  \\
     $k_1$       & $E(2^+_1$) & $E(4^+_1$) & $R_{4/2}$ & $Q(2^+_1$) & $B$(E2;2$^+_1 \rightarrow$ 0$^+_1$) \\  \hline
0.0 & 4.886   &  6.039     & 1.24  & 4.52       & 44.58 \\
0.1 & 4.798   &  6.019     & 1.25  & 6.07       & 48.05 \\
0.2 & 4.654   &  5.979     & 1.28  & 7.79       & 51.52 \\
0.3 & 4.452   &  5.918     & 1.33  & 9.66       & 55.05 \\
0.4 & 4.192   &  5.833     & 1.39  & 11.60       & 58.69 \\
0.5 & 3.875   &  5.721     & 1.47  & 13.52       & 62.73 \\
0.6 & 3.510   &  5.576     & 1.59  & 15.35       & 67.49 \\
0.7 & 3.112   &  5.392     & 1.73  & 17.02       & 73.42 \\
0.8 & 2.705   &  5.165     & 1.91  & 18.49       & 80.88 \\
0.9 & 2.320   &  4.909     & 2.12  & 19.74       & 90.06 \\
1.0 & 1.987   &  4.658     & 2.34  & 18.79       & 81.93 \\  \hline \hline
   $^{24}$Mg,    & $k_2$=1.0  &   &   &  &\\
     $k_1$       & $E(2^+_1$) & $E(4^+_1$) & $R_{4/2}$ & $Q(2^+_1$) & $B$(E2;2$^+_1 \rightarrow$ 0$^+_1$) \\  \hline
0.0&  2.404 &  4.337 &  1.80 &  -7.29 &  27.61  \\
0.1&  2.380 &  4.372 &  1.84 &  -9.56 &  37.65 	\\				
0.2&  2.308 &  4.421 &  1.92 &  -11.81 &  48.48 \\				
0.3&  2.198 &  4.472 &  2.03 &  -13.76 &  58.82 \\				
0.4&  2.067 &  4.504 &  2.18 &  -15.32 &  67.86 \\				
0.5&  1.932 &  4.499 &  2.33 &  -16.51 &  75.50 \\				
0.6&  1.808 &  4.465 &  2.47 &  -17.40 &  81.77 \\				
0.7&  1.702 &  4.423 &  2.60 &  -18.08 &  86.86 \\				
0.8&  1.618 &  4.391 &  2.71 &  -18.58 &  90.91 \\				
0.9&  1.554 &  4.375 &  2.82 &  -18.97 &  94.09 \\				
1.0&  1.509 &  4.378 &  2.90 &  -17.44 &  79.12 \\	\hline	 \hline
   $^{52}$Fe,    & $k_2$=1.0  &   &   &  &\\
     $k_1$       & $E(2^+_1$) & $E(4^+_1$) & $R_{4/2}$ & $Q(2^+_1$) & $B$(E2;2$^+_1 \rightarrow$ 0$^+_1$) \\  \hline
0.0 &	1.020 &	2.295 &	2.25 &	-14.79 &	71.06 \\		
0.1 &	1.020 &	2.299 &	2.25 &	-16.42 &	82.66 \\		
0.2 &	1.015 &	2.306 &	2.27 &	-18.08 &	94.96 \\		
0.3 &	1.006 &	2.316 &	2.30 &	-19.75 &	107.90 \\		
0.4 &	0.993 &	2.327 &	2.34 &	-21.40 &	121.40 \\		
0.5 &	0.978 &	2.342 &	2.39 &	-23.02 &	135.40 \\		
0.6 &	0.960 &	2.359 &	2.46 &	-24.59 &	149.80 \\		
0.7 &	0.941 &	2.378 &	2.53 &	-26.13 &	165.10 \\		
0.8 &	0.923 &	2.399 &	2.60 &	-27.65 &	181.30 \\		
0.9 &	0.903 &	2.420 &	2.68 &	-29.18 &	199.60 \\		
1.0 &	0.883 &	2.434 &	2.76 &	-30.76 &	221.10 \\	\hline	\hline
\end{tabular}
\end{center}
\end{table}

Tables \ref{SiMgFe_k1}-\ref{SiMgFe_k2} display a pattern of correspondence
between the tabulated nuclear observables and the evolution of one of the Hamiltonian parameters,
either $k_1$ or $k_2$, while the other one is kept constant.
In the first case, when $k_1=1$ is constant whereas $k_2$ evolves, the behavior of
the ratio $R_{4/2}$ is very similar for the cases of $^{24}$Mg and $^{52}$Fe: this ratio
first increases reaching a maximum above 3 around $k_2=0.5$ and then decreases for larger values
of $k_2$. The behavior is slightly different for $^{28}$Si having first a minimum at $k_{2}=0.2$
but evolving after that in the same way as in the two previous cases. The absolute energies of
the $2_1^+$ and $4_1^+$ states increase slowly up to the maximum point of $R_{4/2}$ , while after that
the increase of the $2_1^+$ and $4_1^+$ energies is more pronounced.
For the majority of the $k_2$ values,  the ratio $R_{4/2}$ is closer to the rotational limit.
The  reduced transition probabilities $B$(E2;2$^+_1 \rightarrow$ 0$^+_1$) are quite
strong for different values of $k_2$ of the first case, their values
being close to or over 100 ${\rm e}^2 \cdot{\rm fm}^2$. As expected, the ``one unit change" 
matrix elements are to a large extent responsible for the rotational
characteristics, but they still need certain cooperation
of  other matrix elements to create typical characteristics of deformation,
while a too large value of other matrix elements destroys the rotational
features. The discontinuity observed for $^{28}$Si at small
values of $k_2$ is accompanied by a sudden change of the quadrupole moment.

The effects of the $V_2$ part of the interaction with respect to various observables
can be studied using  Table \ref{SiMgFe_k2}, where the parameter $k_2$ is fixed
at the realistic level of 1.0 while  $k_1$ evolves. The dynamics generated by only 
the two-body matrix elements which don't change the occupation number of the subshells 
or induce two-body transitions between the partitions is not capable of creating 
noticeable characteristics of deformation. The increase of $k_1$  drives
a regular decrease of the $2_1^+$ level and a steady growth of the
$R_{4/2}$ ratio, a sign that the deformation trend is under way, though without ever
reaching a pure rotational pattern. A steady  increase is also
observed for the $B$(E2;2$^+_1 \rightarrow$ 0$^+_1$) reduced transition probabilities,
whose values are however lower than 100 ${\rm e}^2\cdot{\rm fm}^2$ in the majority of cases, except
for $^{52}$Fe, whose transition rates $B$(E2;2$^+_1 \rightarrow$ 0$^+_1$) are still
less strong compared to the corresponding cases with $k_1=1.0$ and $k_2$
evolving.

It is expected that the occurrence of rotational motion will increase the low-lying 
level density relative to 
spherical nuclei because of the contribution of emerging rotational bands.
Among the different cases of Tables \ref{SiMgFe_k1} and \ref{SiMgFe_k2} we selected
those that display values of $R_{4/2}$ and $A$ close to rotational and spherical
limits. As can be seen in Table \ref{NOL_SiMgFe_c}, the cases with rotational values
present an enhancement of the level density of the $J=0$ states in the
lowest part of the energy spectrum compared to their spherical
counterparts. These results are almost independent of angular momentum and apply
even for low energies (i.e. the calculation of level density up to 3 MeV
would give qualitatively the same results). The cumulative number of levels (NoL) was
calculated using the moments method. It is convenient for comparison to renormalize
the level density of the moments method making all level densities centered at 1. 
The normalization is achieved by dividing the width of the bin of the original 
Gaussian distribution, which is one, by the mean of the Gaussian found 
using $\frac{N_1*x_1 + N_2*x_2 + \dots}{N_1 + N_2 + \dots}$, 
where $N_i$ is the number of levels in the energy bin, and $x_i$ is the 
mean of the energy bin. In this way all Gaussians get centered at 1. 

This part of the study clarifies the role of the ``one unit change" matrix elements.
The strong presence of the $V_1$ part of the shell-model 
interaction (responsible for mixing of orbitals of the same parity) 
is associated with deformational characteristics 
of the low-lying part of the spectrum. The strong presence of the $V_2$ part of 
the interaction drives the values of all observables away from the rotational limit. 
This situation extends also to the level densities. Not only all rotational cases 
have a larger number of low-energy levels compared to their spherical counterparts, but also they 
are all observed for $k_1=1.0$, while all spherical cases are observed for $k_2=1.0$.

\begin{table}

\caption{Cumulative Number of Levels (NoL) of $J=0$ up to energy 10 MeV
for different ($k_1, k_2$) combinations for $^{28}$Si, $^{24}$Mg and
$^{52}$Fe found with the moments method. The column NoL corresponds 
to the calculation of the moments method, while the column Renorm 
corresponds to the renormalized level density (NoL up to 0.4). }\label{NOL_SiMgFe_c}
\bigskip
\begin{center}
\begin{tabular}{ c  c  c  c  c  c }
\hline \hline
 shape   & case  & nucleus & $R_{4/2}$ & NoL & Renorm\\ \hline
deformed & $k_1=1.0, k_2=0.4$&$^{28}$Si &3.31  & 22  &   60     \\
deformed & $k_1=1.0, k_2=0.5$&$^{28}$Si &3.33  & 17  &   54     \\
deformed & $k_1=1.0, k_2=0.6$&$^{28}$Si &3.21  & 13  &   49     \\
spherical & $k_2=1.0, k_1=0.9$&$^{28}$Si &2.12  & 5  &    34    \\ \hline
deformed & $k_1=1.0, k_2=0.5$&$^{24}$Mg &3.20  & 10  &  24     \\
deformed & $k_1=1.0, k_2=0.6$&$^{24}$Mg &3.21  & 8  &   21    \\
spherical & $k_2=1.0, k_1=0.3$&$^{24}$Mg &2.03  & 6  &    18    \\ \hline
deformed & $k_1=1.0, k_2=0.4$&$^{52}$Fe &3.07  & 236  &  6516  \\
spherical & $k_2=1.0, k_1=0.0$&$^{52}$Fe &2.25  & 30  &  2617    \\  \hline \hline
\end{tabular}
\end{center}
\end{table}

\section{Signatures of a quantum phase transition}

In the previous section we studied the behavior of the two different parts of the 
Hamiltonian, by keeping one part constant and dominant and changing the other. In 
this way we saw that the dominant part gave either rotational ($k_1=1.0$) or 
spherical ($k_2=1.0$) characteristics to the spectrum. In this section we concentrate on
a quantum phase transition that takes place when we change 
simultaneously the strength of the two parts of the Hamiltonian. 

\begin{figure*}
\centering
\includegraphics[height=76mm]{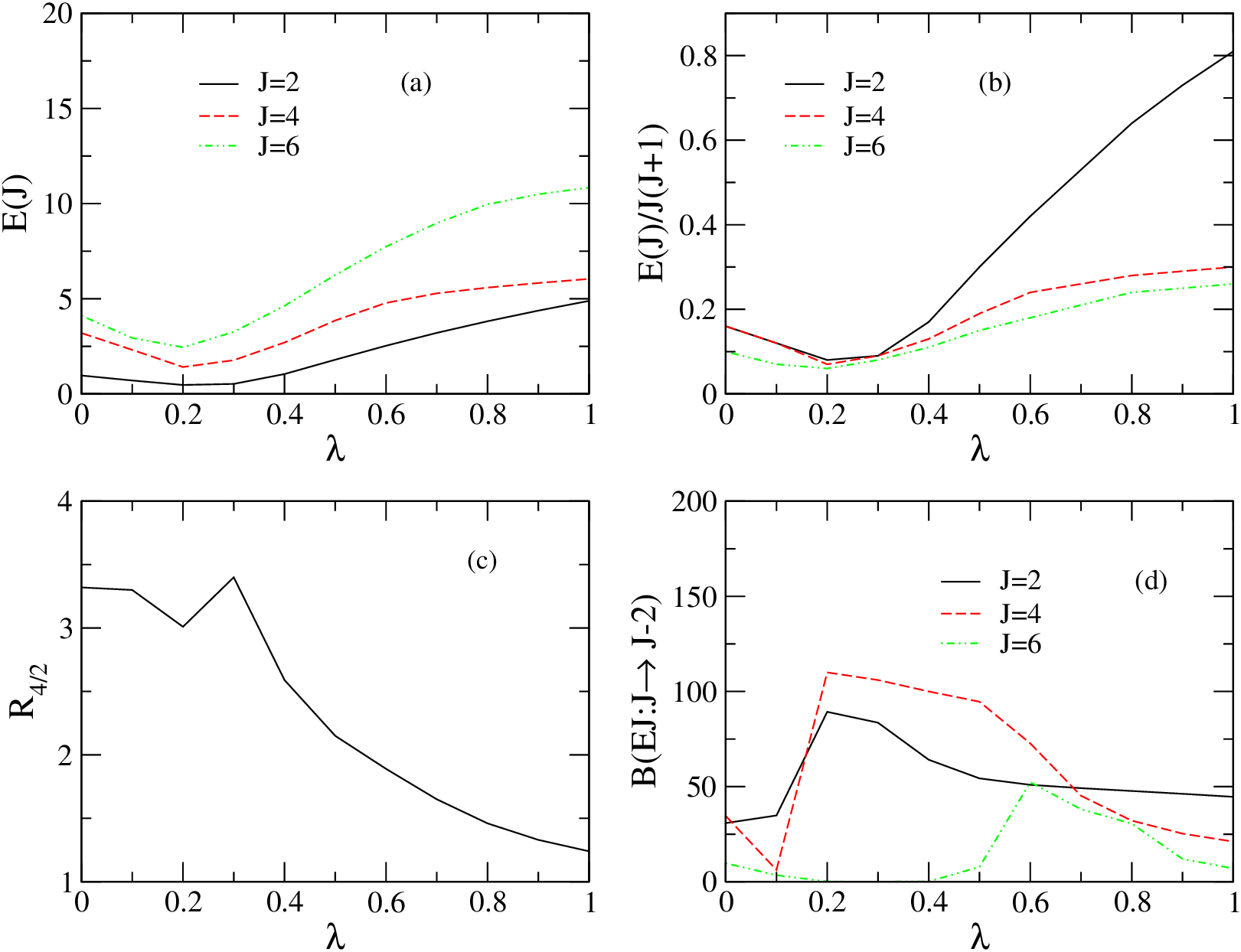}
\caption[]{(a) Yrast 2$^+$, 4$^+$, 6$^+$ energies, (b) ratios $E(J)/J(J+1)$ for $J=0,2,4$,
(c) ratios $R_{4/2}$, (d) electromagnetic transition rates between them
as a function of $\lambda$ for $^{28}$Si.}\label{PT_Si_f}
\end{figure*}

\begin{figure*}
\centering
\includegraphics[height=76mm]{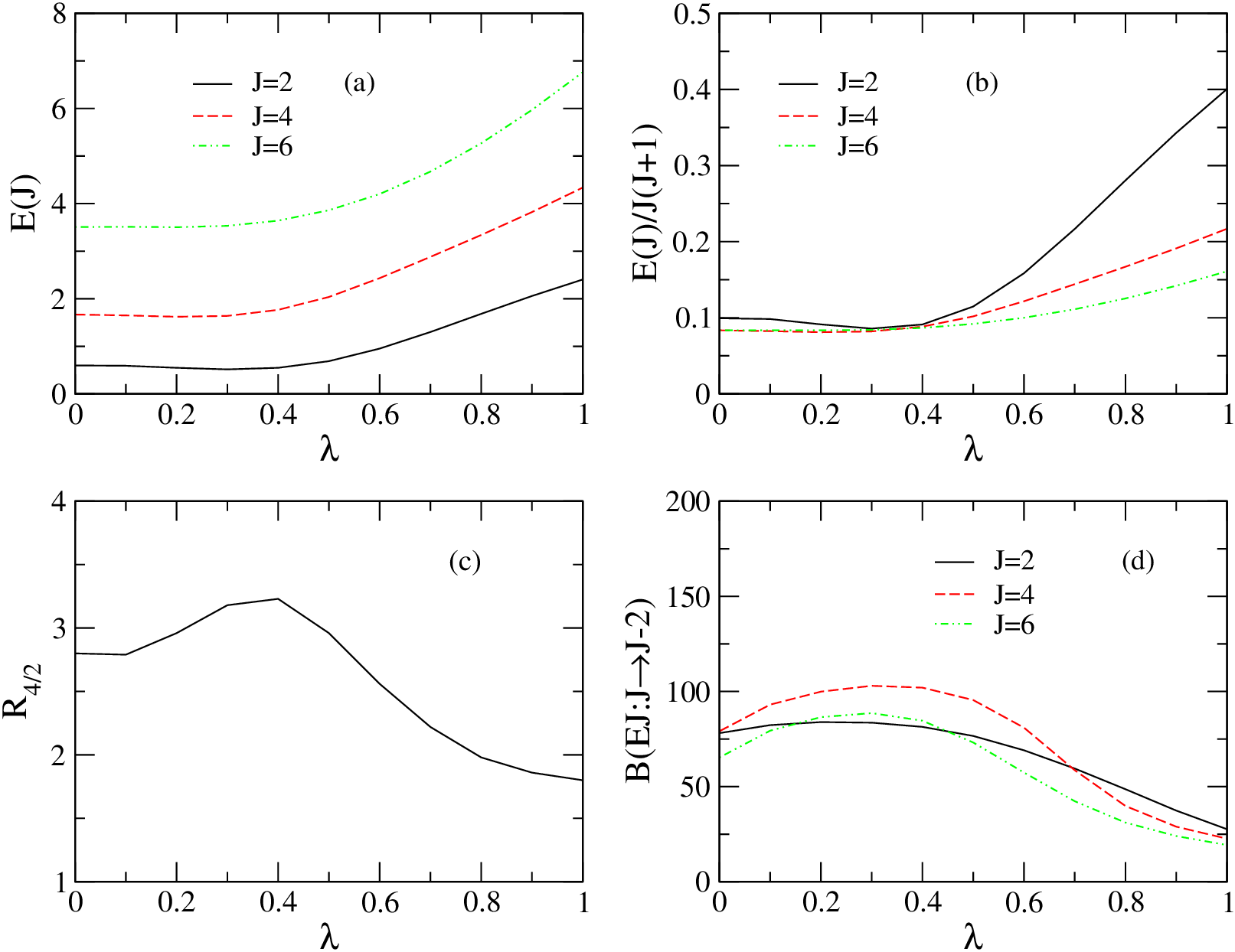}
\caption[]{(a) Yrast 2$^+$, 4$^+$, 6$^+$ energies, (b) ratios $E(J)/J(J+1)$ for $J=0,2,4$,
(c) ratios $R_{4/2}$, (d) electromagnetic transition rates between them
as a function of $\lambda$ for $^{24}$Mg.}\label{PT_Mg_f}
\end{figure*}

\begin{figure*} 
\centering
\includegraphics[height=76mm]{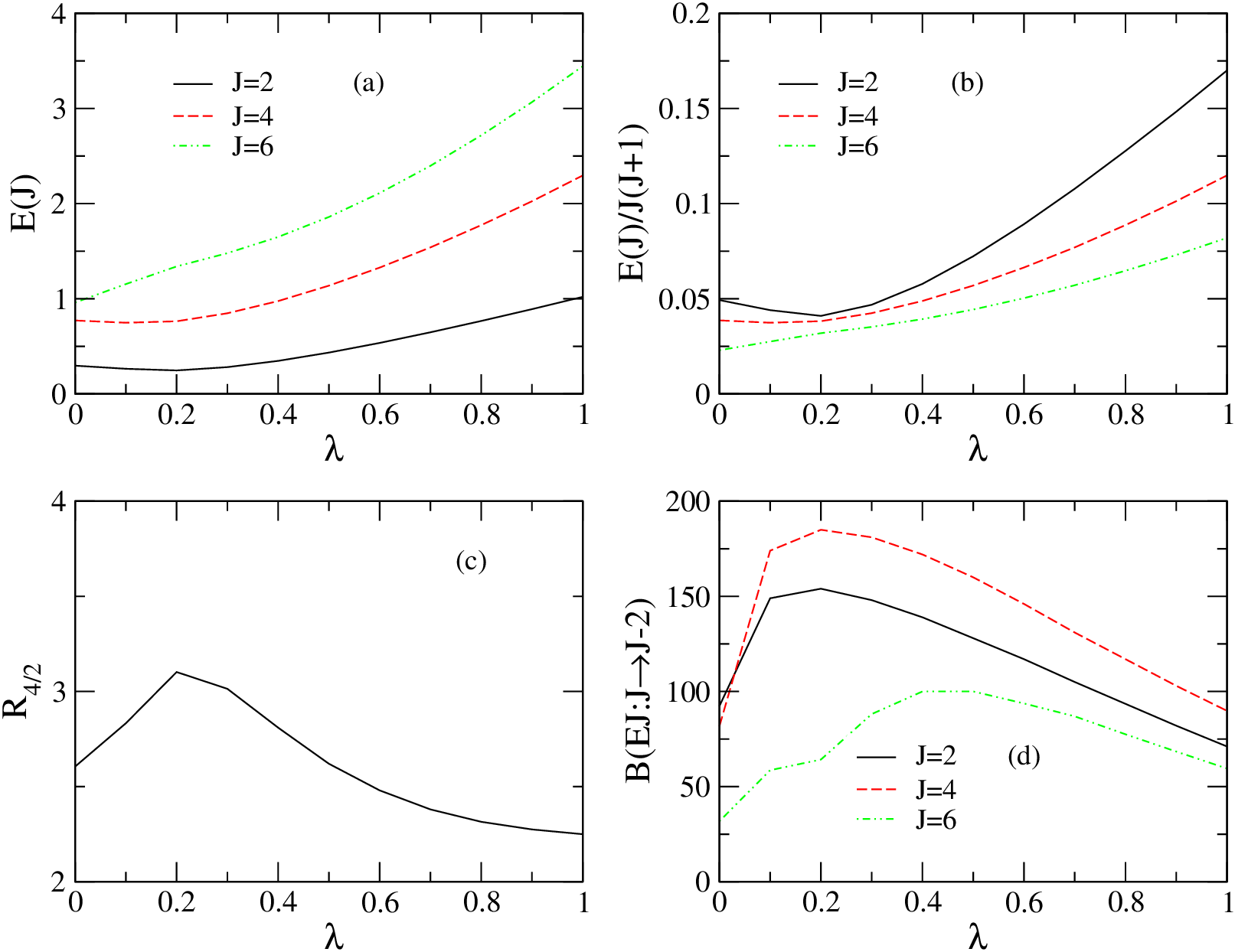}
\caption[]{(a) Yrast 2$^+$, 4$^+$, 6$^+$ energies, (b) ratios $E(J)/J(J+1)$ for $J=0,2,4$,
(c) ratios $R_{4/2}$, (d) electromagnetic transition rates between them
as a function of $\lambda$ for $^{52}$Fe.}\label{PT_Fe_f}
\end{figure*}

Nuclear models have long provided a fertile ground for studying phase transitions
in mesoscopic quantum systems. Quantum phase transitions 
\cite{PT1, PT2, PT3, PT4,PT5,PT6} occur 
when the special observables of a system, called order parameters,
reveal structural, often geometrical, changes as a function
of control quantities. It is convenient to study a quantum phase transition using 
a Hamiltonian of the form   
\begin{equation}
H=h +  (1-\lambda) V_1 +\lambda V_2,  \label{3}
\end{equation}
where the single particle energies part $h$ is fixed, and $\lambda$ is the 
control parameter. In our case
$V_1$ contains the ``one unit change" matrix elements and $V_2$ the rest of matrix 
elements.
By varying $\lambda$ from 0 to 1 in steps 0.1, we study the phase transitional
patterns in the same three nuclei. The results can be found
in Tables \ref{PT_Si}-\ref{PT_Fe} and Figures \ref{PT_Si_f}-\ref{PT_Fe_f}.
We have restricted our study to the yrast states, which exhibit well the effects
of a phase transition.

The $\lambda$-dependence of the low-energy levels presents a minimum at $\lambda$
around 0.2-0.3 for all nuclei and for almost all values of nuclear spin (for $^{24}$Mg,
the minimum of the $2^+_1$ state is displaced to $\lambda=0.3$, while for $^{52}$Fe
the minimum of the $4^+_1$ state is displaced to $\lambda=0.1$). At the same time,
the energy ratio $R_{4/2}$ reaches a maximum, which is always close to a deformed
value, just after, or at, the minimum in the energies of the yrast states. For example, for
$^{24}$Mg the maximum of $R_{4/2}$  appears at $\lambda=0.4$, while for $^{52}$Fe
the maximum  of $R_{4/2}$ and the minimum of the yrast energies coincide. The case
of $^{28}$Si is distinct from the other two, since its quadrupole moment changes
abruptly from negative to positive values and the $R_{4/2}$ ratio has two maxima
for different types of deformation. The second maximal $R_{4/2}$ value appears
for $\lambda=0.3$. Another quantity that reflects the effects of the phase
transition is the quadrupole moment of the $2^{+}$ state that has a minimum close to the
values of $\lambda$ where other observables have their extremal values.

We note that the ratio $E(J)/J(J+1)$ (effective inverse
moment of inertia) is almost independent of $J$ for all nuclei, from $\lambda=0.0$
up to the value of $\lambda$ where the energy ratio $R_{4/2}$ has its maximum
for each particular nucleus. The reduced transition probabilities are also
sensitive to the phase transition, showing a maximum close to the point of
minimum energy of the yrast states. The  probabilities
$B$(E2;$6^+_1\rightarrow4^+_1)$ for $^{28}$Si and $^{52}$Fe  have a
maximum for slightly  greater values of $\lambda$.

\begin{table*}

\caption{Yrast energies of $2^+$, $4^+$ and $6^+$ (MeV), ratios $R_{4/2}$, quadrupole
moments $Q(2^+_1)$ (${\rm e}\cdot {\rm fm}^2$) and
reduced transition probabilities $B$(E2;2$^+_1 \rightarrow$ 0$^+_1$), $B$(E2;4$^+_1 \rightarrow$ 2$^+_1$),
$B$(E2;6$^+_1 \rightarrow$ 4$^+_1$) (${\rm e}^2\cdot{\rm  fm}^2$) for $^{28}$Si}   \label{PT_Si}
\bigskip
\begin{center}
\begin{tabular}{ r  r  r  r  r  r  c  c  c }
\hline \hline
$\lambda$& $2^+_1$ & $4^+_1$ & $6^+_1$ & $R_{4/2}$ & $Q(2^+_1)$& B(E2;2$^+_1 \rightarrow$ 0$^+_1$) & B(E2;4$^+_1 \rightarrow$ 2$^+_1$) & B(E2;6$^+_1 \rightarrow$ 4$^+_1$) \\ \hline
0.0 &	0.964 &	3.197 &	4.110 &	3.32 & -11.5&	30.75	&	34.61	&	9.70  \\
0.1 &	0.702 &	2.314 &	2.938 &	3.30 & -12.14&  34.80	&	6.22	&	3.46  \\
0.2 &	0.469 &	1.410 &	2.447 &	3.01 & 18.74&	89.30	&	109.60	&	0.10  \\
0.3 &	0.521 &	1.771 &	3.257 &	3.40 &  18.63&  83.56	&	105.80	&	0.06  \\
0.4 &	1.041 &	2.699 &	4.624 &	2.59 & 17.36&	64.08	&	100.40	&	0.16  \\
0.5 &	1.793 &	3.857 &	6.252 &	2.15 &15.42&	54.32	&	94.61	&	7.77  \\
0.6 &	2.529 &	4.777 &	7.739 &	1.89 &13.1&	50.94	&	72.65	&	52.32 \\
0.7 &	3.203 &	5.280 &	8.973 &	1.65 &10.7&	49.24	&	45.16	&	38.08 \\
0.8 &	3.815 &	5.587 &	9.959 &	1.46 &8.41&	47.74	&	32.07	&	30.52 \\
0.9 &	4.373 &	5.826 &	10.492 &1.33 &6.35&	46.21	&	25.30	&	11.96 \\
1.0 &	4.886 &	6.039 &	10.842 &1.24 &4.52&     44.58	&	21.07	&	6.97  \\ \hline \hline
\end{tabular}
\end{center}
\end{table*}

\begin{table*}

\caption{Yrast energies of $2^+$, $4^+$ and $6^+$ (MeV), ratios $R_{4/2}$, quadrupole
moments $Q(2^+_1)$ (${\rm e}\cdot{\rm fm}^2$) and reduced transition probabilities $B$(E2;2$^+_1 \rightarrow$ 0$^+_1$), $B$(E2;4$^+_1 \rightarrow$ 2$^+_1$),
$B$(E2;6$^+_1 \rightarrow$ 4$^+_1$) (${\rm e}^2\cdot{\rm fm}^2$) for $^{24}$Mg} \label{PT_Mg}
\bigskip
\begin{center}
\begin{tabular}{ r  r  r  r  r  r  c  c  c }
\hline \hline
$\lambda$& $2^+_1$ & $4^+_1$ & $6^+_1$ & $R_{4/2}$ & $Q(2^+_1)$ & B(E2;2$^+_1 \rightarrow$ 0$^+_1$) & B(E2;4$^+_1 \rightarrow$ 2$^+_1$) & B(E2;6$^+_1 \rightarrow$ 4$^+_1$) \\ \hline
0.0 &	0.596 &	1.667 &	3.507 &	2.80 &	-16.32 &	78.09	&	79.09	&	65.32 \\
0.1 &	0.590 &	1.649 &	3.512 &	2.79 &	-18.02 &	82.34	&	93.06	&	79.36 \\
0.2 &	0.548 &	1.620 &	3.504 &	2.96 &	-18.65 &	83.89	&	99.93	&	86.48 \\
0.3 &	0.515 &	1.640 &	3.533 &	3.18 &	-18.87 &	83.64	&	102.70	&	88.61 \\
0.4 &	0.547 &	1.766 &	3.642 &	3.23 &	-18.68 &	81.37	&	101.50	&	84.58 \\
0.5 &	0.688 &	2.036 &	3.860 &	2.96 &	-17.97 &	76.60	&	95.45	&	73.21 \\
0.6 &	0.951 &	2.433 &	4.201 &	2.56 &	-16.65 &	69.12	&	81.00	&	57.41 \\
0.7 &	1.300 &	2.881 &	4.672 &	2.22 &	-14.71 &	59.51	&	58.57	&	42.26 \\
0.8 &	1.683 &	3.339 &	5.267 &	1.98 &	-12.28 &	48.55	&	39.82	&	31.08 \\
0.9 &	2.058 &	3.821 &	5.970 &	1.86 &	-9.68 &	        37.37	&	28.94	&	23.99 \\
1.0 &	2.404 &	4.337 &	6.761 &	1.80 &	-7.29 &	        27.61	&	22.63	&	19.30 \\ \hline \hline

\end{tabular}
\end{center}
\end{table*}

\begin{table*}

\caption{Yrast energies of $2^+$, $4^+$ and $6^+$ (MeV) states, ratios $R_{4/2}$, quadrupole
moments $Q(2^+_1)$ (${\rm e}\cdot {\rm fm}^2$) and reduced transition probabilities $B$(E2;2$^+_1 \rightarrow$ 0$^+_1$), $B$(E2;4$^+_1 \rightarrow$ 2$^+_1$),
$B$(E2;6$^+_1 \rightarrow$ 4$^+_1$) (${\rm e}^2\cdot{\rm fm}^2$) for $^{52}$Fe} \label{PT_Fe}
\bigskip
\begin{center}
\begin{tabular}{ r  r  r  r  r  r  c  c  c }
\hline \hline
$\lambda$& $2^+_1$ & $4^+_1$ & $6^+_1$ & $R_{4/2}$ & $Q(2^+_1)$&  B(E2;2$^+_1 \rightarrow$ 0$^+_1$) & B(E2;4$^+_1 \rightarrow$ 2$^+_1$) & B(E2;6$^+_1 \rightarrow$ 4$^+_1$) \\ \hline

0.0 &	0.296 &	0.771 &	0.960 &	2.61 &	-20.68 &        92.28	&	81.66	&	31.76  \\
0.1 &	0.264 &	0.748 &	1.154 &	2.83 &	-25.27 &	149.20	&	173.90	&	58.55  \\
0.2 &	0.246 &	0.763 &	1.340 &	3.10 &	-25.60 &	154.10	&	184.90	&	64.16  \\
0.3 &	0.281 &	0.847 &	1.479 &	3.01 &	-25.02 &	148.00	&	181.30	&	88.09  \\
0.4 &	0.347 &	0.975 &	1.649 &	2.81 &	-24.00 &	138.50	&	171.80	&	100.30 \\
0.5 &	0.434 &	1.137 &	1.860 &	2.62 &	-22.70 &	127.90	&	159.50	&	100.10 \\
0.6 &	0.535 &	1.327 &	2.111 &	2.48 &	-21.24 &	116.60	&	145.80	&	93.69  \\
0.7 &	0.647 &	1.540 &	2.397 &	2.38 &	-19.68 &	105.10	&	131.40	&	86.89  \\
0.8 &	0.766 &	1.773 &	2.717 &	2.32 &	-18.06 &	93.48	&	117.10	&	77.49  \\
0.9 &	0.890 &	2.025 &	3.067 &	2.28 &	-16.42 &	82.01	&	103.20	&	68.40  \\
1.0 &	1.020 &	2.295 &	3.444 &	2.25 &	-14.79 &	71.06	&	89.81	&	59.53  \\ \hline \hline

\end{tabular}
\end{center}
\end{table*}

\begin{figure*}
\centering
\includegraphics[height=78mm]{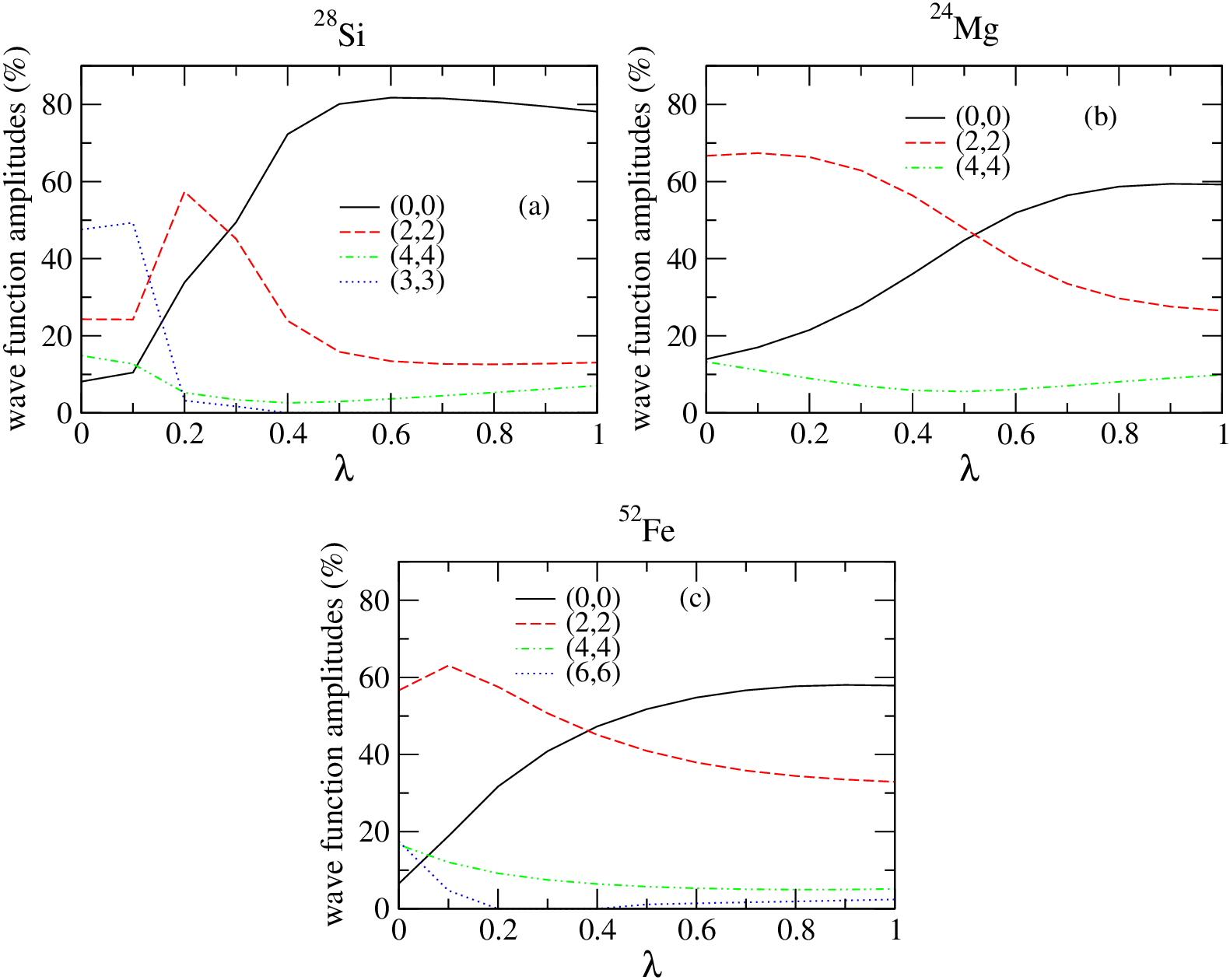}
\caption[]{Amplitudes of the ground state wave function expanded in terms of the
proton and  neutron components coupled to angular momenta
(0,0), (2,2), (3,3), (4,4), (6,6) as a function of $\lambda$ for (a) $^{28}$Si,
(b) $^{24}$Mg, and (c) $^{52}$Fe.}\label{SiMgFe_wvf}
\end{figure*}

The ground state wave function 
also displays the signs
of a quantum phase transition. Fig. \ref{SiMgFe_wvf} shows the amplitudes of this
function expanded in terms of single-particle orbitals for protons
and neutrons coupled to angular momenta $(J_{n},J_{p})=$(0,0), (2,2), (3,3), (4,4), (6,6) as a
function of $\lambda$. The results for the amplitudes of the couplings (0,0), (2,2)
and (4,4) are quite similar for $^{24}$Mg and $^{52}$Fe. The (0,0) coupled pairs
have their minimum amplitudes for small values of $\lambda$ while the (2,2) coupled
pairs are stronger for the same values of $\lambda$. Basically, up until the point
of the quantum phase transition, the (2,2) coupled pairs are the strongest
components of the ground state wave function, a behavior consistent with deformation characteristics.
After the critical point, their amplitudes fall down and the amplitudes of the (0,0)
coupled pairs rise, becoming eventually the strongest components of the
wave function, a typical feature of the vibrational limit. The amplitudes of the (4,4)
coupled pairs have their largest values for the smallest $\lambda$ and then they
slowly decrease, taking an almost steady value after the point of the phase transition.
The behavior of the amplitudes of $^{28}$Si for $\lambda=0.0$ and 0.1 differs 
from other nuclei, as the amplitudes have a steady but still coherent behavior
with the (2,2) component being stronger than the (0,0) one, but with a clear predominance
of the (3,3) component over all others. This steady behavior suddenly breaks
for $\lambda=0.2$, with the components moving abruptly to the values they would have
if the quadrupole moments had had a steady sign following the behavior of the
amplitudes in other nuclei. For $^{52}$Fe, up to the transitional point, the
(6,6) component seems to be also of some importance.

These results suggest that a nuclear system governed by the Hamiltonian (\ref{3})
undergoes a phase transition at $\lambda=0.2$, with the rotational
characteristics being more evident for $\lambda \leq 0.2$ and declining for
$\lambda >0.2$. One might expect that close to the transition point, where the
excitation energies have their minimum values, an enhancement of the level density would
be observed, at least at relatively low energy. Previous studies in the
framework of the IBM model for large boson numbers have confirmed this
enhancement \cite{Cejnar} in the spectrum of $0^+$ states. Enhancement in the
number of low-lying $0^+$ states has also been observed experimentally
\cite{Meyer}  in the rare earth region for the transitional nucleus $^{154}$Gd.

\begin{figure}
\centering
\includegraphics[height=103mm]{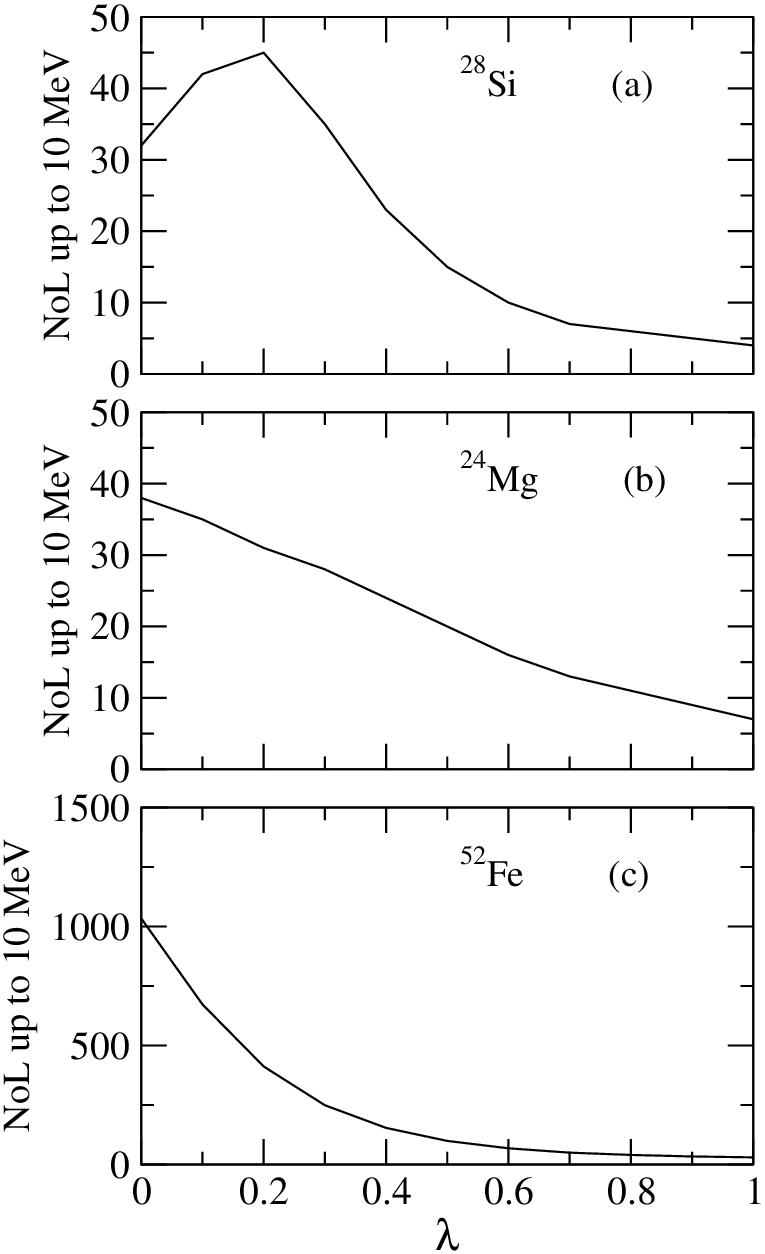}
\caption[]{Number of levels up to 10  MeV as a function of $\lambda$ for (a) $^{28}$Si,
(b) $^{24}$Mg, (c) $^{52}$Fe.}\label{NoL_SiMgFe_f}
\end{figure}

In order to search for signs of the collective enhancement, we calculate the number
of $0^+$ states up to 10 MeV for selected three nuclei at
different  values of the parameter $\lambda$, as shown in
Table \ref{NoL_SiMgFe_L0_T} and Figure \ref{NoL_SiMgFe_f}. These results are
qualitatively independent of the  angular momentum used $-$ different spins
show the same behavior of the level density. The number of levels was
calculated using the moments method as well as its renormalized version when
all level densities are centered at unity. 

No signs of collective enhancement are
observed just at the transitional point. In all cases there is a sharp drop at
the number of levels as a function of $\lambda$. This result doesn't change
even if we use a smaller energy interval to calculate the number
of levels, for instance up to 3 MeV. For $^{28}$Si, a peak appears for $\lambda=0.2$,
i.e. at the point of the transition, however this peak has to  be attributed to
the sudden change of the quadrupole moment at $\lambda=0.2$, since there is no
similar effect in other two nuclei, whose quadrupole moment has a steady sign. 
The vicinity of the phase transition point that in a finite system is always
smeared as a crossover can be studied in more detail by means of the invariant
correlational entropy \cite{volya}.

\begin{table}

\caption{Cumulative Number of Levels (NoL) with $J = 0$ up to energy 10 MeV for different values
of $\lambda$ for $^{28}$Si, $^{24}$Mg and $^{52}$Fe. The column NoL corresponds
to the calculation of the moments method, while the column Renorm corresponds
to the renormalized level density (NoL up to 0.4 MeV).}\label{NoL_SiMgFe_L0_T}
\bigskip
\begin{center}
\begin{tabular}{ r  r  c  r  r  r  c  r  r  r  c }
\hline	\hline
$^{28}$Si&          &  & &$^{24}$Mg &       &  &  &$^{52}$Fe&	& \\ 	
$\lambda$&	NoL & Renorm & & $\lambda$&	NoL & Renorm &  &$\lambda$&	NoL&  Renorm\\ \hline
0.0 & 32 & 67 &  & 0.0 & 38 & 44 & & 0.0 & 1034 & 12853 \\
0.1 & 42 & 75 &  & 0.1 & 35 & 40 & & 0.1 & 673 & 10435 \\
0.2 & 45 & 75 &  & 0.2 & 31 & 36 & & 0.2 & 412 & 8278  \\
0.3 & 35 & 63 &  & 0.3 & 28 & 33 & & 0.3 & 249 & 6581 \\
0.4 & 23 & 49 &  & 0.4 & 24 & 31 & & 0.4 & 154 & 5284 \\
0.5 & 15 & 38 &  & 0.5 & 20 & 28 & & 0.5 & 99 & 4354 \\
0.6 & 10 & 31 &  & 0.6 & 16 & 26 & & 0.6 & 68 & 3746 \\
0.7 & 7 & 27 &  & 0.7 & 13 & 24 & & 0.7 & 50 & 3248 \\
0.8 & 6 & 24 &	& 0.8 & 11 & 22 & & 0.8 & 40 & 2942 \\
0.9 & 5 & 23 &  & 0.9 & 9 & 20  & & 0.9 & 34 & 2731 \\
1.0 & 4 & 22 &	& 1.0 &	7 & 19 & & 1.0 & 30 & 2617 \\ \hline
\end{tabular}
\end{center}
\end{table}

Last, among the different level densities calculated,  we selected those few ones that
indicate a spherical or a deformed shape, according to their  $R_{4/2}$ value.
According to  Table  \ref{NOL_SiMgFe_com} where we collected the   results,  deformed
cases always  have enhanced level density compared to the spherical cases. This
seems to be a general  result consistently observed among all the cases studied.

\begin{figure}[h]
\centering
\includegraphics[height=76 mm]{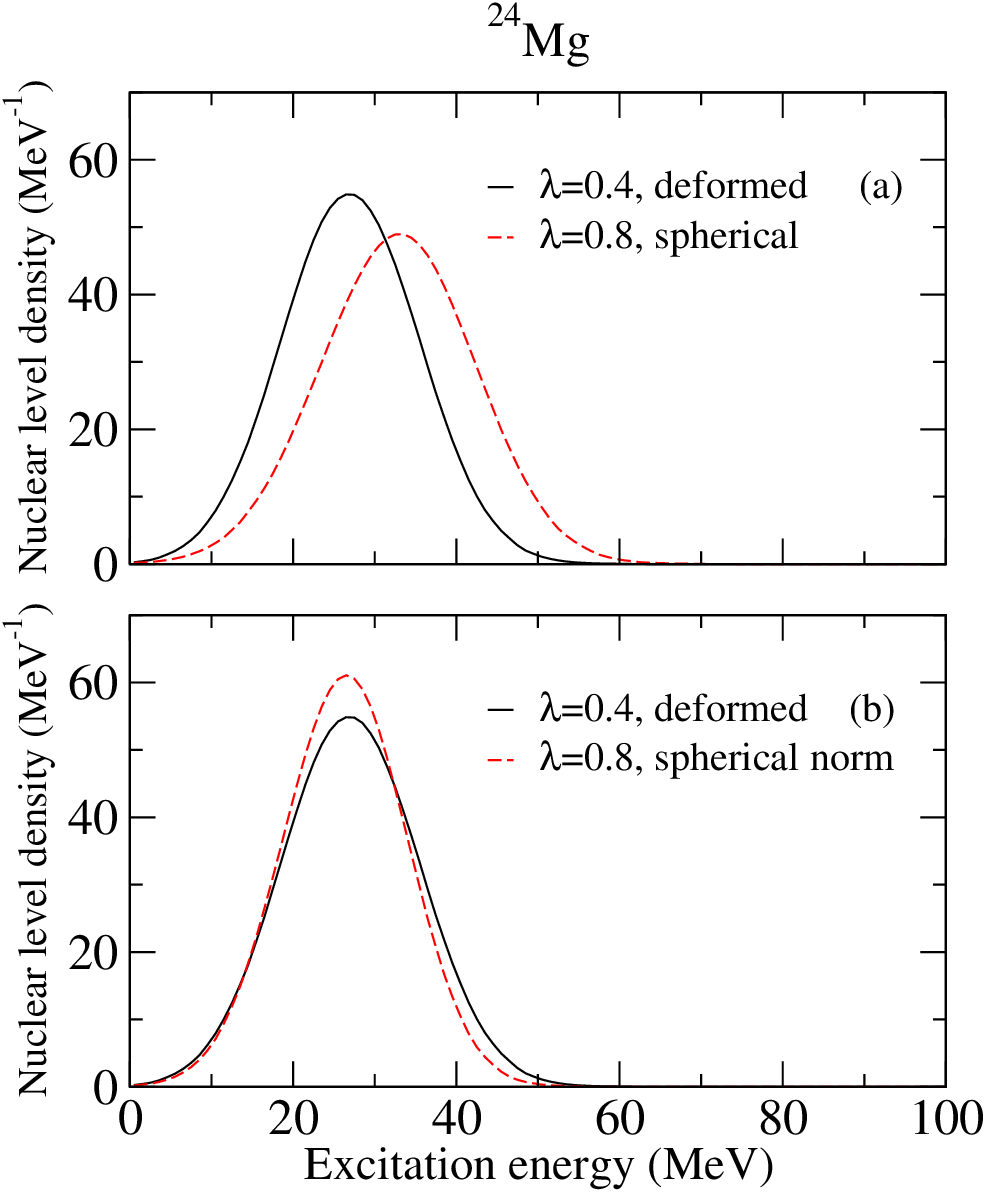}
\caption[]{Nuclear level densities for $J=0$, for $^{24}$Mg and for the points given in Table \ref{NOL_SiMgFe_com}. 
The upper panel, (a), corresponds to the level density as given by the moments method. 
The lower panel, (b), shows the level density that corresponds to the spherical case, 
normalised to the width of the deformed level density.}\label{defvsSphMg}
\end{figure}

\begin{table}
\caption{Cumulative Number of Levels (NoL) up to 10 MeV energy for spherical or deformed cases
which appear for various values of $\lambda$ for $^{28}$Si, $^{24}$Mg, and  $^{52}$Fe
nuclei}\label{NOL_SiMgFe_com}
\bigskip
\begin{center}
\begin{tabular}{ r  r  r  r  r }
\hline \hline
 shape   & case      &nucleus          & $R_{4/2}$ & NoL \\  \hline
deformed & $\lambda$=0.0& $^{28}$Si &3.32  & 32          \\
spherical & $\lambda$=0.5& $^{28}$Si & 2.15  & 15         \\ \hline
deformed & $\lambda$=0.4& $^{24}$Mg &3.23  & 24          \\
spherical & $\lambda$=0.8& $^{24}$Mg &1.98  & 11          \\ \hline
deformed & $\lambda$=0.2& $^{52}$Fe &3.10  & 412          \\
spherical & $\lambda$=1.0& $^{52}$Fe &2.25  & 30         \\  \hline \hline
\end{tabular}
\end{center}
\end{table}

\section{Discussion}

In this study, the fixed Hilbert space of the shell model has been probed by varying the numerical parameters 
of the Hamiltonian while keeping intact all exact conservation laws.
This allows us to study the evolution of physical observables and the corresponding
level density. Technically the shell-model Hamiltonian was divided in two parts. 
The part $V_1$ included the two-body matrix elements which induce the transfer of 
one nucleon between the partitions, whereas the part $V_2$ contained
remaining matrix elements. By varying the strength of the two parts of the
Hamiltonian, we have followed the changes of the energy spectrum, quadrupole moments and
transition probabilities for selected nuclei in the $sd$ and $pf$ shells.
The results confirm that the ``one unit change" matrix elements are 
responsible for the appearance of rotational characteristics, lowering energy of the $2^+_1$
and $4^+_1$ levels, inducing the $R_{4/2}$ values typical for a rotor and large reduced
transition probabilities between rotational states. On the other hand, the $V_2$ part of the interaction breaks the
rotational characteristics and induces a vibrational behavior.

Collective modes in nuclei strongly influence the level density at the
low-energy part of the spectrum, the phenomenologically known effect called the {\it collective enhancement}. 
By selecting the rotational and vibrational cases resulting from the variation of the shell
model Hamiltonian, one can clearly see that the deformed nuclear spectra are richer in
low-lying levels compared to the spherical ones, a clear
indication of collective enhancement. The enhancement has to be compensated at higher energy
unless we extend our orbital space; for the fixed space, the compensation occurs beyond 
the borderline of applicability of the used shell-model version.

The role of the ``one unit change" matrix elements as the carriers of deformation 
is so pronounced that one can even  see  a quantum phase transition between deformed  
and spherical (often with soft vibrations as predecessors of the shape instability) 
phases of the system, by simultaneously varying the $V_1$ and $V_2$ parts
of the Hamiltonian. The phase transition reveals itself in the ground state wave function 
of the system, the energy spectrum and transition probabilities.
No similar phase transition has been observed by dividing the Hamiltonian in other
combinations. The phase transition reveals itself by the cooperative dynamical action of many 
components of the interaction present in the shell-model Hamiltonian. 

Our preliminary results have shown that 
in the case in odd-odd nuclei the ``one unit change" matrix elements 
affect noticeably the whole energy spectrum. For odd-odd nuclei we would expect to 
see signs of collective enhancement even at the transitional point. 

\section*{Acknowledgments}
The work was supported by the NSF grant PHY-1404442.
We are thankful to B.A. Brown and R.A. Sen'kov for numerous discussions; V.Z. acknowledges
useful discussions with N. Auerbach during the visit to Tel Aviv supported by the Binational
Science Foundation US-Israel.

\end{document}